\documentclass[conference]{IEEEtran}
\IEEEoverridecommandlockouts
\usepackage{cite}
\usepackage{amsmath,amssymb,amsfonts}
\usepackage{algorithmic}
\usepackage{graphicx}
\usepackage{textcomp}
\usepackage{xcolor}
\usepackage{float} 
\usepackage{orcidlink}
 
\def\BibTeX{{\rm B\kern-.05em{\sc i\kern-.025em b}\kern-.08em
    T\kern-.1667em\lower.7ex\hbox{E}\kern-.125emX}}
\usepackage{fancyhdr}
\usepackage{lastpage}
\usepackage[english]{babel}
\usepackage{tikz}
\usetikzlibrary{calc}
\usepackage{lipsum}
\begin{document}

\title{A Deep Transfer Learning Approach on Identifying Glitch Wave-form in Gravitational Wave Data \\
}

\author{\IEEEauthorblockN{Reymond Mesuga\IEEEauthorrefmark{1} \orcidlink{0000-0002-8180-2245} and Brian James Bayanay }
\IEEEauthorblockA{\textit{Department of Physical Sciences, College of Science} \\
\textit{Polytechnic University of the Philippines, Manila, Philippines} \\
\IEEEauthorrefmark{1}Corresponding author: rrmesuga@iskolarngbayan.pup.edu.ph}

}
\maketitle
\begin{abstract}
LIGO interferometer is considered the most sensitive and complicated gravitational experimental equipment ever built. Its main objective is to detect the gravitational wave from the strongest events in the universe by observing if the length of its 4-kilometer arms change by a distance 10,000 times smaller than the diameter of a proton. Due to its sensitivity, interferometer is prone to the disturbance of external noises which affects the data being collected to detect the gravitational wave. These noises are commonly called by the gravitational-wave community as glitches. This study focuses on identifying those glitches using different deep transfer learning algorithms. The extensive experiment shows that algorithm with architecture VGG19 recorded the highest AUC-ROC among other experimented algorithm with 0.9898. While all of the experimented algorithm achieved a considerably high AUC-ROC, some of the algorithm suffered from class imbalance of the dataset which has a detrimental effect when identifying other classes.
\end{abstract}


\section{\textbf{Introduction}}
\subsection{\textbf{Physics Background}}
Gravitational waves (GWs) are deformations in spacetime that result from astrophysical phenomena involving celestial objects of masses much heavier than that of the sun moving at speeds up to a significant fraction of the speed of light, mainly called compact objects. GWs result from either mergers of binaries of compact objects, such as binary black hole (BBH) mergers, binary neutron star (BNS) mergers, neutron star-black hole binaries, white dwarf binaries, etc., or from self-production by a massive release of energy from astrophysical phenomena like stellar collapse (i.e., supernovae).  

Ever since the first direct detection of GWs by the joint collaboration of LIGO and Virgo on September 14, 2015, the field of gravitational-wave astronomy has become one of the rising fields of research in contemporary physics, and with upgrades to the LIGO detectors in the US as well the VIRGO detector in Italy, combined with the newly-operational KAGRA Observatory in Japan and the operation of the LISA Mission in future years, more and more GW events are being and will be detected, and with these detections come terabytes of data that are in great need of accurate analysis, to ensure that the signals that these observatories are indeed signals from outer space and not noise, either of terrestrial or electromagnetic origin. To build a somewhat good foundation of how the GW event data is being gathered, it is a must to give a short elaboration on the experimental setup of these observatories.

In the case of LIGO-Hanford and LIGO-Livingston, based in the United States, the 2 observatories are ground-based Michelson interferometers with arms spanning 4 km, where a 20-watt laser is fired, passing through a power recycling mirror, which then fully transmits light incident from the laser and reflects light from the other side increasing the power of the light field between the mirror and the subsequent beam splitter. From the beam splitter, the light travels along two orthogonal arms,
and by using partially reflecting mirrors, Fabry–Pérot cavities are created in both arms that increase the effective path length of laser light in the arm. When a GW of sufficient energy passes through the interferometer, the space-time in the local area is deformed, manifested through the effective change in length of one or both Fabry–Pérot cavities. This change in length will cause the light in the cavity to be slightly out of phase with the incoming light, which will lead to the cavity/s being out of coherence, and the laser light, which are tuned to destructively interfere at the detectors, will have a slightly periodically varying detuning, resulting in a measurable signal, with the detectors’ sensitivities up to lengths 10000 times smaller than the diameter of the proton \cite{1}. Due to this and the LIGO-US
detectors being ground-based, factors such as instrument noise and environmental influence \cite{2} to name a few, the LIGO detectors not only records the GW strain data, but also over 200,000 auxiliary channels that monitor instrument behavior and environmental conditions \cite{3}. Then, the GW strain data and the data from the auxiliary channels (which may or may not contain legitimate GW strain data) are then subject to data analysis.

\begin{figure*}[h!]
\includegraphics[width=\textwidth,height=9cm, width=18cm]{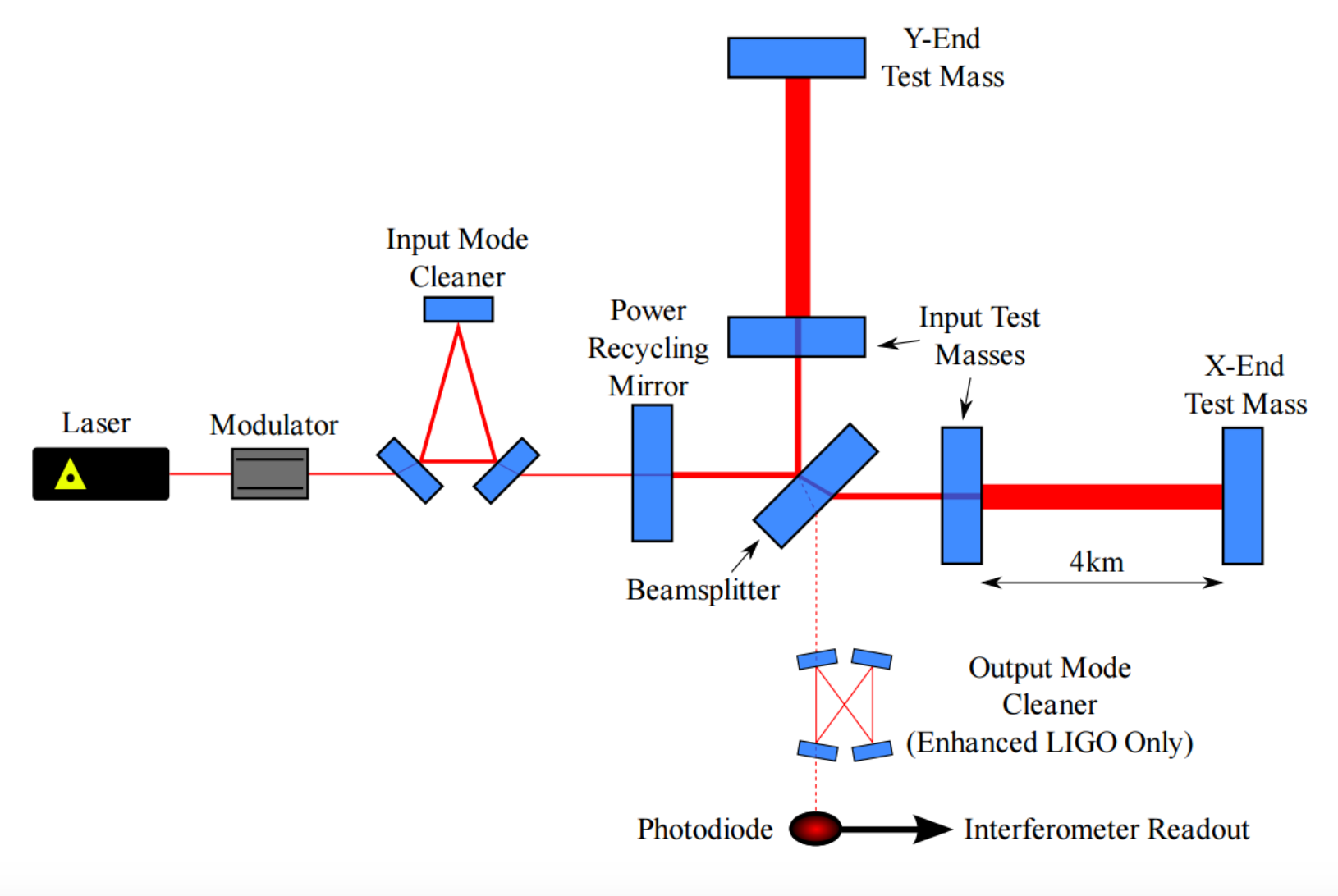}
\caption{Experimental setup of the Advanced LIGO Detector,
\cite{1}}
\label{fig:1}
\end{figure*}

In the analysis of GW data, of particular concern are transient, non-Gaussian noise features, called glitches, which are instrumental or environmental in nature (caused by e.g., small ground motions, ringing of the test-mass suspension system at
resonant frequencies, or fluctuations in the laser) and come in a wide variety of time-frequency amplitude morphologies \cite{4}, and can mimic true gravitational wave signals and can hinder
sensitivity conditions \cite{5}. These glitches are then classified by common origin and/or similar morphological characteristics \cite{4}.
\subsection{\textbf{Deep Learning}}
\label{subsec:B}
The most used method used in the identification and classification of glitches in GW data is by means of machine learning (ML) algorithms \cite{5, 6, 7, 8, 9, 10, 11} such as dictionary learning \cite{5}, similarity learning \cite{6}, deep transfer learning \cite{7} among many a method. Building on this, the LIGO-Virgo Collaboration has advocated citizen science involvements in the search for these glitches, e.g., Gravity Spy \cite{1}, which speeds up the process of refining the increasingly large amount of GW data.

\begin{figure}[h!]
\includegraphics[width=0.9999\columnwidth]{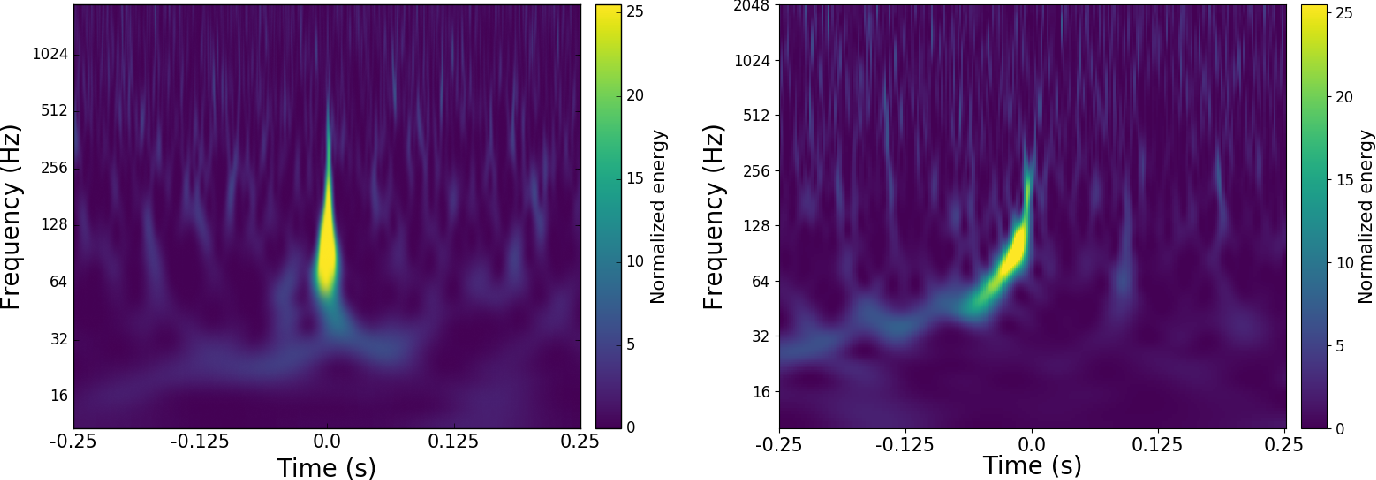}
\caption{Comparison of images from a glitch signal (left) and a GW signal (right) in LIGO O2 data,
\cite{26}}
\label{fig:2}
\end{figure}

The refining of GW data is of paramount importance to the scientific community, particularly to physicists working in general relativity, cosmology, astrophysics, quantum gravity, etc. because the phenomenon itself serves as a validity of theories of gravity, mainly general relativity, and opens new questions to various fields in physics, like the mass threshold at which GW should occur, one of particular interest is the most recent detection (O3) of LIGO-Virgo, GW190814 \cite{12}, which indicates a “mass gap” between the heaviest neutron stars and the lightest black holes. These GW data is also used in the investigation of the Hubble expansion of the universe \cite{13, 14}, cosmic inflation via the existence of a stochastic gravitational-wave background \cite{15, 16}, existence of dark matter \cite{17}, among many new physics. With that being said, the efficiency of the different machine learning algorithms
used in searching, classifying. and mitigating glitches should be considered, in order to obtain the most accurate data possible, and for this purpose, this experiment aims to compare some of the different
machine learning algorithms used in noise detection in GW data, and their respective efficiencies in doing so.

\begin{figure}[h!]
\includegraphics[width=0.9999\columnwidth]{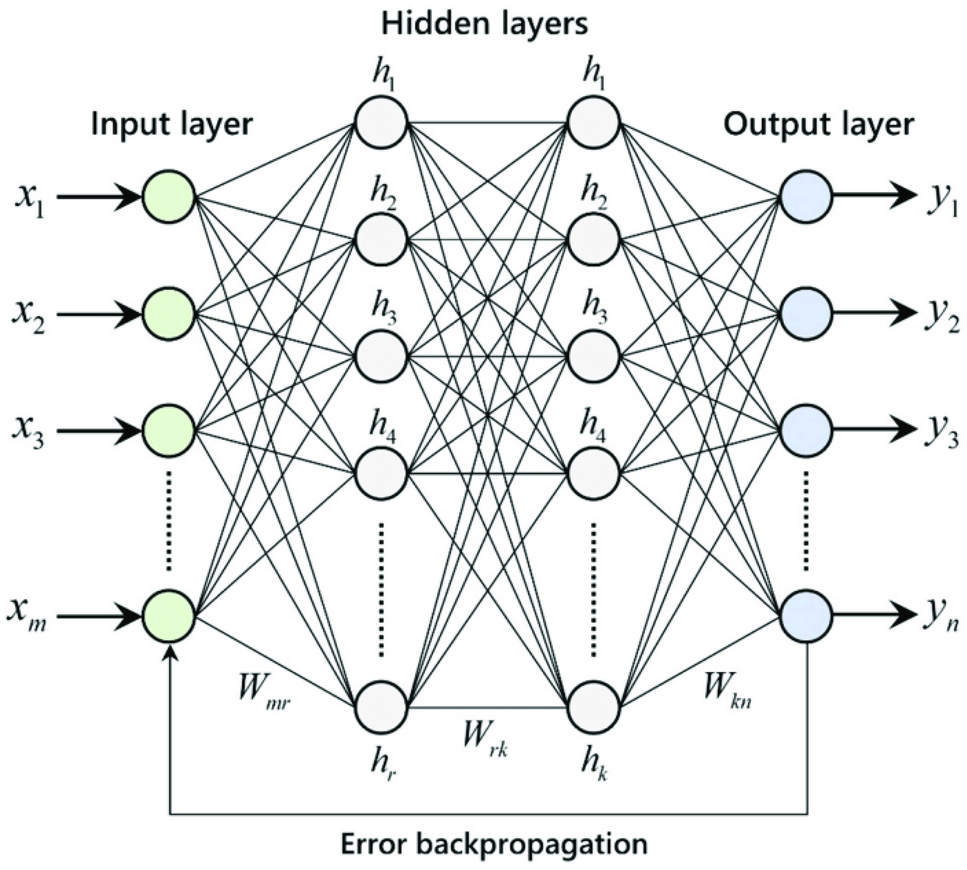}
\caption{ A simple diagram of artificial neural network,
\cite{27}}
\label{fig:3}
\end{figure}

The use of deep learning algorithms will be adopted in this paper. Deep learning (DL) is a type of machine learning (ML) algorithm where it uses a so called artificial neural network (ANN) to learn from different input data (i.e., images, sounds and texts) \cite{18, 19}. ANN was inspired on how the human brain works. In fact, ANN has its own version of neuron that functions almost similar to biological neuron called artificial neuron which represents the nodes that can be found in the hidden layer (see Fig. \ref{fig:3}). In a biological neuron, if the signals of information received by the synapse are strong enough (or surpass a certain threshold) \cite{19}. When it comes to artificial neuron, if the weight of an input is not enough then the neuron will not be activated. In an image classification problem, each pixels on the image will serve as the inputs in input layer (see Fig. \ref{fig:3}).

After setting the pixels of an image as inputs, the ANN will then apply a randomly selected weights and multiply each on inputs. This can be defined as: 
\begin{equation}
\label{eq:1}
y = w*x + b
\end{equation}

where $y$ is the output, $w$ is the weight, $x$ is the input and $b$ is a special kind of weight called bias. This function is also the equation of the line.

Unfortunately, ANN (or/and DL algorithms in general) won’t give any meaningful results just by just using a linear transformation. The application of non-linearity is needed because inputs such as images,
sounds and texts are naturally non-linear. To apply non-linearity, each neuron on the hidden layer needs a so called activation function. Now, the output $y$ on each neuron can be defined as:
\begin{equation}
\label{eq:2}
y=\varphi(x*w+b)
\end{equation}

where $\varphi$ is the activation function. Note that equation \ref{eq:2} only represents an output $y$ using a single
input $x$. In general, output $y$ of a neuron is the summation of all input $x$ with their corresponding weights and bias and can be defined as equation \ref{eq:3}:
\begin{equation}
\label{eq:3}
y_k=\varphi(\sum_{i=0}^{m} x_{i}*w_{ki}+b_{k})
\end{equation}

\begin{equation}
\label{eq:4}
\sum_{i=0}^{m} x_{i}*w_{ki}+b_{k}
\end{equation}

The most common and up-to-date activation function is the ReLU activation function. This can be defined as:

\begin{equation}
\label{eq:5}
f (x)=max(0 , x)
\end{equation}
The idea here is that, if the value of equation \ref{eq:4} is less than or equal to zero, the output $y$ will be automatically set as zero and will deactivated, Otherwise, if it is greater than zero, then the output $y$ will stay as it is and hence activated \cite{20}.

Now, one cannot expect a deep learning algorithm to have an accurate and meaningful result just by using a random selected weights. In order for a deep learning algorithm to have a more accurate result, the algorithm need to be trained. Training the algorithm involves the adjustments of the weights. These weights need to be adjusted in a way that it fits on what is needed by the algorithm to give more accurate result. These adjustments will continue as long as the algorithm has not reached yet its minimum loss. The measurement of loss will be done using a \textit{Loss Function} which measures how good or bad a deep learning is to classify each classes on the dataset. The most common and up-to-date loss function used in a multi-class image classification is the \textit{Categorical Cross Entropy Loss} \cite{22}. Meanwhile, the most common and up-to-date optimization algorithm that is used to adjust the weights is called the \textit{Adam Optimizer} \cite{21}. This is the simple approach to understand how artificial neural network (ANN) works in an image classification. That is, by setting the the pixels of an image as inputs, applying random weights on each inputs, applying non-linear transformation on each inputs with their corresponding weights using activation function, training the model to lower the loss measured by a loss function and adjusting the weights using the optimizer. 

\begin{figure}[h!]
\includegraphics[width=0.9999\columnwidth]{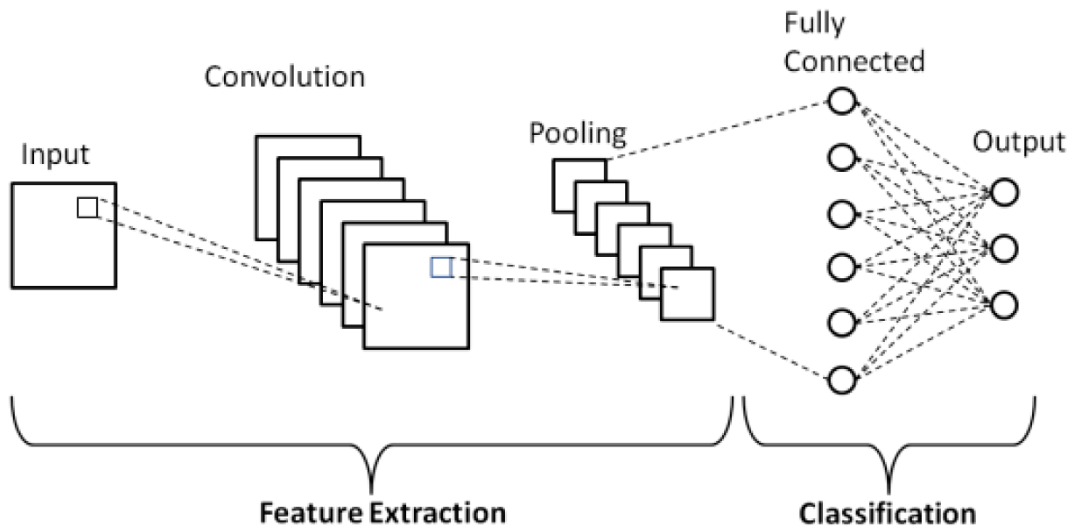}
\caption{ A simple diagram of convolutional neural network (CNN),
\cite{23}}
\label{fig:4}
\end{figure} 

Artificial neural network (ANN) has many variations of algorithms. The most successful ANN algorithm is called Convolutional Neural Network (also called as CNN or ConvNet) \cite{23,24}. The structure of
CNN can be divided into two parts which the the base and the head. The base of CNN is used to extract the features from an image and is formed primarily of three basic components namely \textit{convolution layer}, \textit{ReLU activation function} and \textit{maximum pooling}. On the other hand, the head of the CNN is responsible for determining the class of the image. The main usage of convolution layer is to filter an
image for a particular feature. Meanwhile, ReLU activation detects the feature within the filtered image and maximum pooling is responsible for the enhancement \cite{24}. 

The general objective of the study is to evaluate the performance of various deep transfer learning models in glitch waveform detection in gravitational-wave data. The specific objectives are as follows:
a) Evaluate and Compare the performance of different algorithms using specific evaluation metric, b) Identify how the quantity of the data affects the performance of the algorithm, and c) Identify how the complexity of the algorithm affects its performance in identifying glitch wave-forms.

\section{\textbf{Methodology}}

\subsection{\textbf{Data Gathering and Preparation}}

The dataset used in this study were gathered from a Kaggle repository that have been classified as a part of the Gravitational Spy Dataset \cite{28}. The only difference between the Kaggle and original version of dataset is that image data found in a Kaggle repository has no axes and was already divided into train set, validation set and test set which contains 22348 (70\%), 4800 (15\%), and 4720 (15\%) respectively with the total of 31868 image data. Table \ref{tab:1} shows the number of data per classes.  As observed, the class Blip
contains the majority of images which has 1821 images and the classes 1400Ripples,
None\_of\_the\_Above, Chirp, Air\_Compressor, Wandering\_Line and Paired\_Doves did not even make
it above a hundred.

\begin{table}[h!]
    \caption{No. of Data per Classes}
    \label{tab:1}
    \centering
    \begin{tabular}{|c|c|}
    \hline
    Glitch Classes & No. of Data \\
    \hline
        Blip            &       1821 \\
        Koi\_Fish           &     706 \\
        Low\_Frequency\_Burst  &   621 \\
        Light\_Modulation    &    512 \\
        Power\_Line          &    449 \\
        Extremely\_Loud       &   447 \\
        Low\_Frequency\_Lines  &   447 \\
        Scattered\_Light     &    443 \\
        Violin\_Mode        &     412 \\
        Scratchy           &     337 \\
        1080Lines           &    328 \\
        Whistle              &   299 \\
        Helix                &   279 \\
        Repeating\_Blips       &  263 \\
        No\_Glitch           &    150 \\
        Tomte       &            103 \\
        None\_of\_the\_Above   &     81 \\
        1400Ripples     &         81 \\
        Chirp            &        60 \\
        Air\_Compressor    &       58 \\
        Wandering\_Line     &      42 \\
        Paired\_Doves        &     27 \\
    \hline
    \end{tabular}
\end{table}

\subsection{\textbf{Deep Transfer Learning Algorithms}}
The type of deep learning algorithms that will be used in the experiment are all deep transfer learning algorithm. Deep transfer learning or simply transfer learning uses pre-trained architectures as its base
algorithm. As mentioned in section \ref{subsec:B}, the structure or architecture of convolutional neural network (CNN) can be divided into two parts (i.e., base and head). Most transfer learning algorithm uses CNN architecture as well. The only difference is that, the base of transfer learning algorithms were trained already using different images in the past. The most notable dataset that is commonly used to train a base for a pre-trained algorithm is called ImageNet which contains 1.2 million images that has 1000 different classes \cite{25}. The architectures tested were ResNet \cite{29}, VGG \cite{30}, Xception \cite{31}, Inception \cite{32}, and DenseNet \cite{33}. To be specific, the following pre-trained architectures used are as  follows: ResNet101, ResNet101V2, ResNet152, ResNet50, ResNet50v2, VGG16, VGG19, Xception, InceptionResnetV2, and DenseNet169.

\subsection{\textbf{Performance Metrics}}

\begin{figure}[h!]
\includegraphics[width=0.9999\columnwidth]{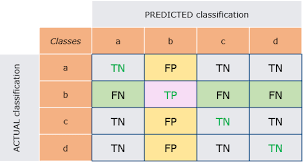}
\caption{ Confusion matrix for multi-classes,
\cite{34}}
\label{fig:5}
\end{figure}

\begin{figure}[h!]
\includegraphics[width=0.999\columnwidth]{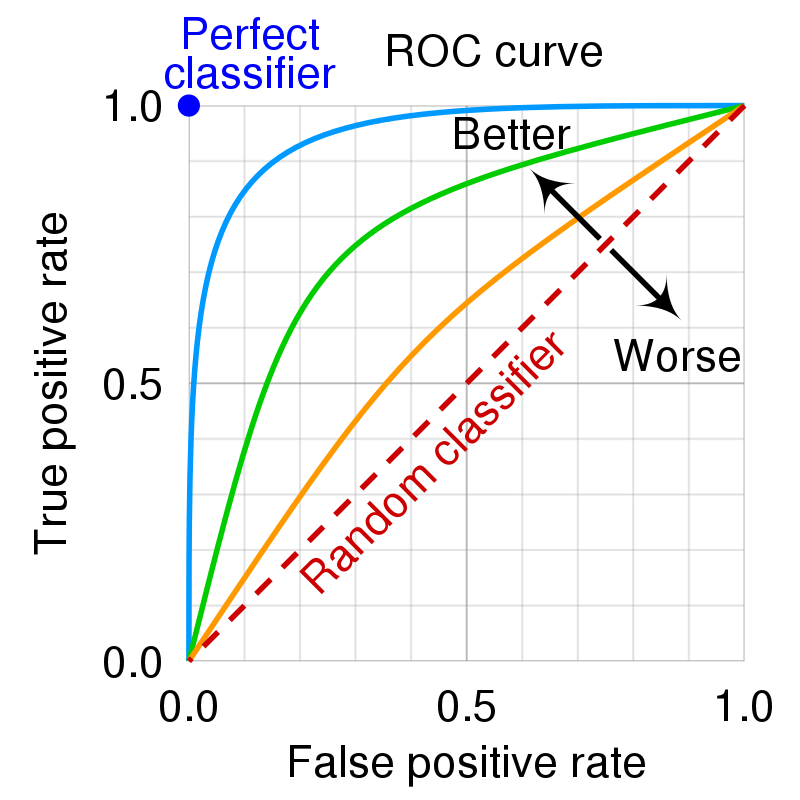}
\caption{ AUC-ROC,
\cite{35}}
\label{fig:6}
\end{figure}

 The below figures (i.e., Fig. \ref{fig:5} and \ref{fig:6}) shows the illustration for confusion matrix and AUC-ROC will be discussed further in the following sub-subsection.
 
\subsubsection{\textbf{Confusion Matrix}}
Confusion matrix for multi-class classification works the same way as in binary classification. It can be visualize in Fig. \ref{fig:5} where $y-axis$ corresponds to the actual classification of samples while $x-axis$ corresponds to the predicted classification. 

The main goal of using a confusion matrix is to visualize the quantity of samples that are correctly and wrongly predicted by the classifier. The samples in confusion matrix can be categorized into four types namely \textit{True Positive} (TP), \textit{True Negatives} (TP), \textit{False Positives} (FP), and \textit{False Negatives} (FN).  

\subsubsection{\textbf{Area Under the Curve of Receiver Operating Characteristic (AUC-ROC)}}

AUC-ROC is primarily used in binary classification problems classification according to their one-vs-all precision-recall-curves. AUC-ROC is primarily used to see if the classifier was able to separate the classes in the data. In order to plot the actual ROC, \textit{True Positive Rate} (TPR) and \textit{False Positive Rate} (FPR) must be computed with many different threshold and plotted as in Fig. \ref{fig:6} where $x-axis$ corresponds the the FPR while $y-axis$ corresponds to TPR. The ideal value of AUC-ROC is 1 which means that the classifier was able to separate perfectly the classes in the data. The lowest is 0.5 which means the classifier is just making predictions randomly (see Figure \ref{fig:6}. The AUC-ROC metric was used as the primary performance metric in this study as it is commonly used in field of deep learning and is better than using accuracy for decision making. 

\section{\textbf{Discussion of Results}}
\begin{figure}[h!]
\includegraphics[width=0.9999\columnwidth]{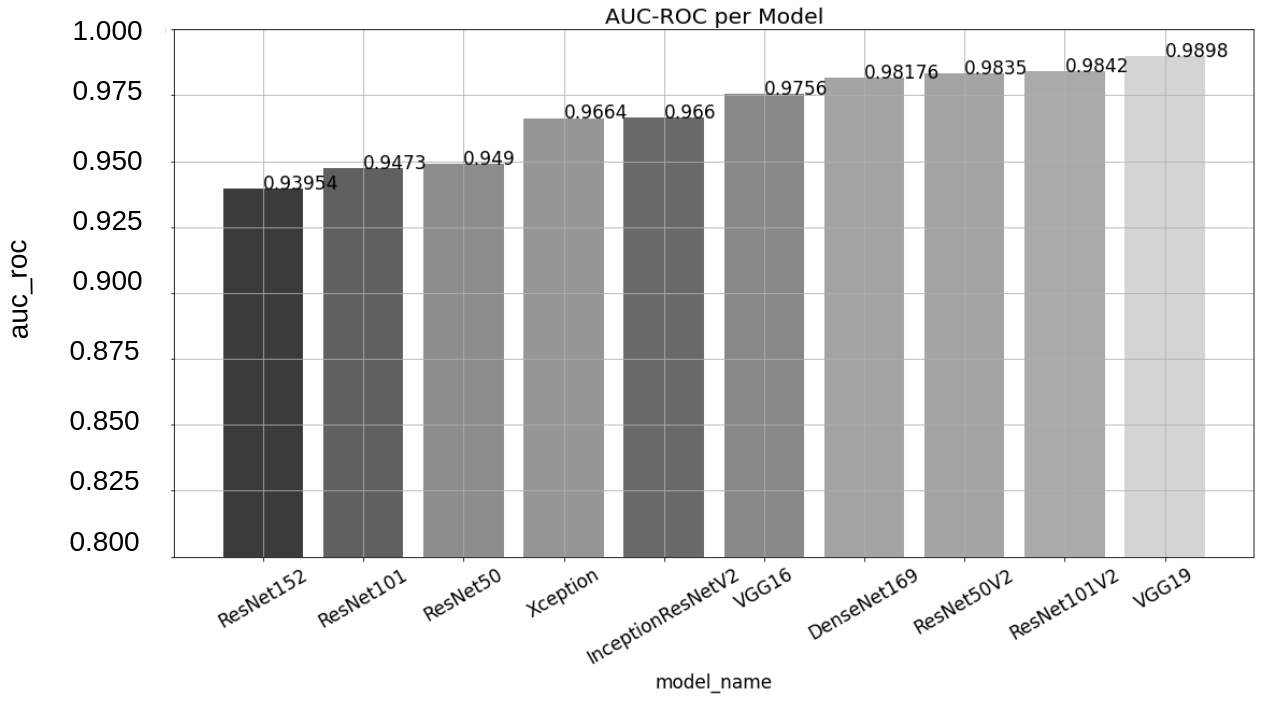}
\caption{ AUC-ROC per algorithm}
\label{fig:7}
\end{figure}

The bar plot in Fig. \ref{fig:7} shows that VGG19 recorded the highest AUC-ROC overall followed by ResNet101V2, ReNet50V2, and DenseNet169 all of which achieved an AUC-ROC higher than 0.98. Meanwhile, ResNet152 recoded the lowest AUC-ROC of 0.93954 which is not bad. Overall, all the algorithm have recorded an AUC-ROC higher than 0.90 which is considerably high or decent.

\begin{figure}[h!]
\includegraphics[width=0.9999\columnwidth]{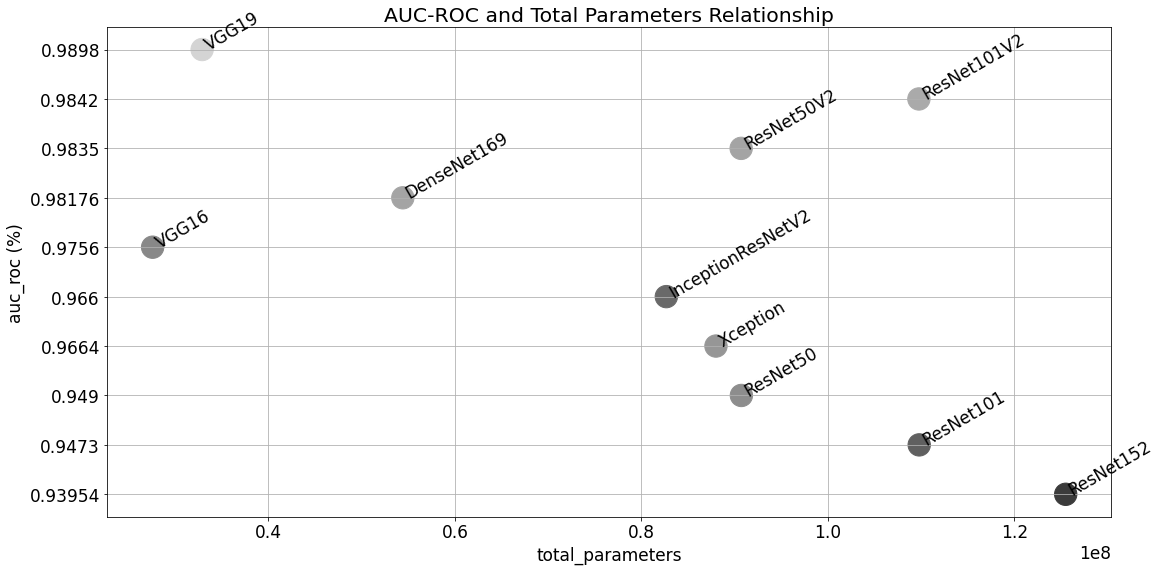}
\caption{ AUC-ROC vs Complexity}
\label{fig:8}
\end{figure}

To better understand the difference in performance between each algorithm, it is also good to plot their AUC-ROC vs Complexity. The term complexity here means the total parameters of the model which can be interpreted as how complex they are compared to others. 

As shown in Fig. \ref{fig:8}, lighter algorithm architectures like VGG16 and VGG19 outperformed several architectures like InceptionResNetV2, Xception, ResNet50, ResNet101, and ResNet52 despite having a more complex or bigger network.

\begin{figure}[h!]
\includegraphics[width=0.9999\columnwidth]{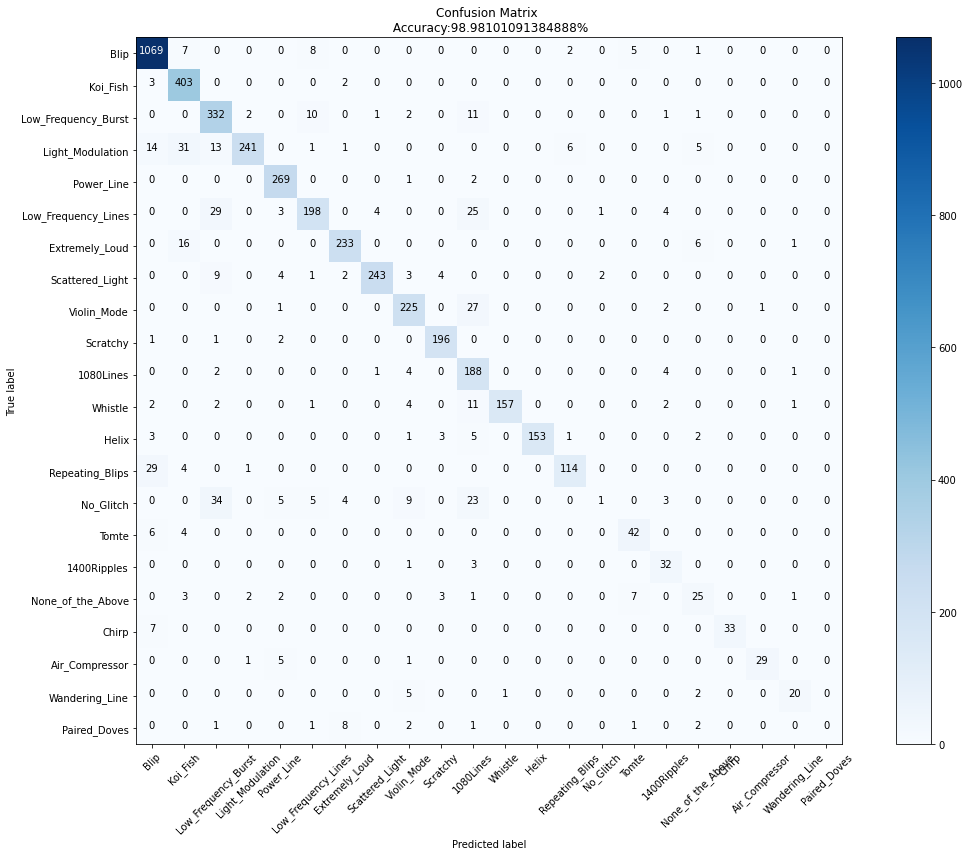}
\caption{ Confusion matrix of VGG19 (AUC-ROC=0.9898)}
\label{fig:9}
\end{figure}

Fig. \ref{fig:9} shows the confusion matrix of VGG19. It is visible that it is capable of identifying low quantity classes such as the class Air\_Compressor and Wandering\_Line. Note that image data in Gravity Spy Dataset has a problem in class imbalanced (see Table \ref{tab:1}). The two mentioned classes are one of the classes with lowest number of data which greatly affect the performance of other algorithms. Despite that, VGG19 still able to identify those classes at considerably decent performance.

\begin{figure}[h!]
\includegraphics[width=0.9999\columnwidth]{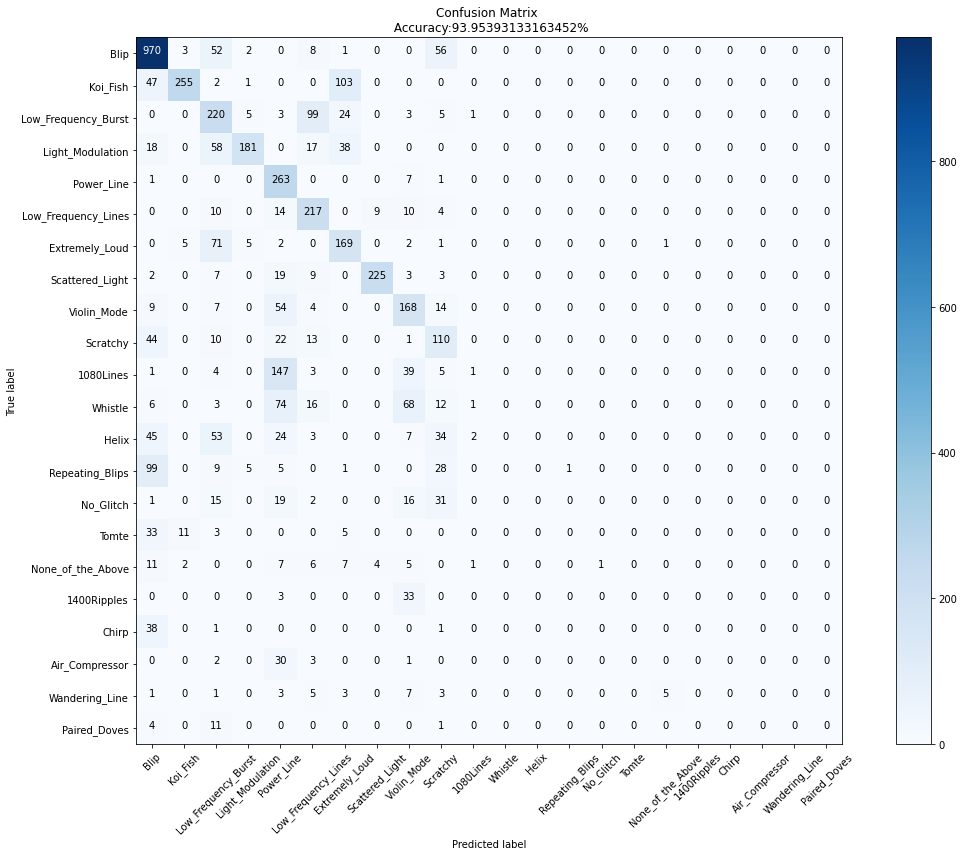}
\caption{ Confusion matrix of ResNet152 (AUC-ROC=0.93954)}
\label{fig:10}
\end{figure}

Meanwhile, Fig. \ref{fig:10} shows the confusion matrix of ResNet152 which achieved an AUC-ROC of 0.93954 which is the lowest recorded AUC-ROC among other experimental algorithms. It was visible that ResNet152 were not able to properly almost half of the classes in the Gravity Spy Dataset. Despite that, it still achieved a considerably high AUC-ROC due to the fact that the upper half classes in the confusion matrix contains majority of the samples in the dataset. 

With that, it could be interpreted that in order to use deep transfer learning algorithms in a dataset that contains many classes with imbalanced samples, AUC-ROC have to be higher as much as possible with, at least, 0.97 or above in order to perform decently in the real world. 

\section{\textbf{Conclusion}}
There are still room for improvements when it comes to the domain of deep transfer learning applications in identifying glitch wave-form in gravitational wave data. The need for improvement is needed especially on the part of producing more data to train algorithms. Using a decent quantity of data with balanced classes is really a necessity and will definitely help the performance of the algorithm as it is one of the major problems found in this study. All of the algorithms featured in this study shows a promising performance that could potentially useful for the use of gravitational-wave scientists. But, there are still differences in performances between algorithms in which some perform better than the other. The differences on the performances of each algorithm should one of the major factor when applying deep transfer learning in production specifically on identifying glitch wave-forms in gravitational wave data.

\section{\textbf{Acknowledgement}}
The authors would like to thank Mr. Mark Anthony Burgonio, MS for allowing this study to be conducted under his Advance Laboratory course.

\section{\textbf{CRediT Authorship Contribution Statement}}
\textbf{Reymond Mesuga:} Conceptualization, Methodology, Software, Investigation, Validation, Formal Analysis, Resources, Data Curation, Writing-Original Draft, Writing-Review \& Editing, Visualization, Project Administration, Supervision; \textbf{Brian James Bayanay} Investigation, Writing-Original Draft, Writing-Review and Editing; 

\section{\textbf{Conflict of Interest}}
The authors declare that they have no known competing financial interests or personal relationships that
could have appeared to influence the work reported in this paper.

\vspace{12pt}

\end{document}